\documentclass[twocolumn,prl,floats,showpacs,preprintnumbers,amsmath,amssymb]{revtex4}
\usepackage{hyperref}

\usepackage{amsfonts}
\usepackage{amssymb}
\usepackage[dvips]{graphicx}
\usepackage{latexsym}

\usepackage{dcolumn}
\usepackage{bm}

 \newcommand{\beq}{\begin{equation}}
 \newcommand{\eeq}{\end{equation}}
 \newcommand{\bea}{\begin{eqnarray}}
 \newcommand{\eea}{\end{eqnarray}}
 \newcommand{\lpart}{\raise.3ex\hbox{$\stackrel{\leftarrow}{\partial}$}}
 \newcommand{\rpart}{\raise.3ex\hbox{$\stackrel{\rightarrow}{\partial}$}}
 \newcommand{\ldr}{\raise.3ex\hbox{$\stackrel{\leftarrow}{\delta^r}$}}

\begin{document}

\preprint{HIP-2004-24/TH}

\title{A CMB/Dark Energy Cosmic Duality}

\author{Kari Enqvist}
\email{kari.enqvist@helsinki.fi}
\author{Martin S. Sloth \footnote{Address from Sept. 2004:
    Department of Physics, University of California, Davis, CA 95616,
    USA.}}
\email{sloth@physics.ucdavis.edu}

\affiliation{Department of Physical Sciences, University of Helsinki, and
Helsinki Institute of Physics \\
P.O. Box 64, 00014 University of Helsinki, Finland}

\pacs{98.80.Cq,98.80.-k}

\begin{abstract}
We investigate a possible connection
 between the suppression of the power at low multipoles in the CMB
 spectrum and the late time acceleration.
 We show that, assuming a cosmic IR/UV duality between
the UV cutoff and a global infrared cutoff given by the
 size of the future event horizon,
 the equation of state of the dark energy can be related to the apparent
 cutoff in the CMB spectrum.
 The present limits on the equation of state of dark energy are shown
 to imply an IR cutoff in the CMB multipole interval of $9>l>8.5$.
\end{abstract}

\maketitle

While the $\Lambda CDM$ model describes the CMB angular spectrum
well, WMAP data \cite{WMAP} indicates that there could be a
problem at largest scales. Compared with the model predictions, at
low multipoles there appears to be a lack of power; according to
simulations, the observed WMAP quadrupole has a low probability
\cite{WMAP}. Of course, the explanation might simply be
cosmic variance combined with bad luck, but there is also the
possibility that the low multipoles signal for new physics.
Effectively, reducing the angular power at low multipoles requires
an introduction of an cutoff in the infrared. Such cutoff could be
dynamical \cite{Contaldi:2003zv} and due to, say, properties of
the inflaton potential such that the slope of the potential
changes much during the beginning of the last 65 e-folds. A period
during which the inflaton kinetic energy dominates could also be
one possible explanation. The common feature of these scenarios is
that they rely on local physics, and that tuning of the parameters
is necessary.

Another, although admittedly a more speculative possibility, is
that there exist global constraints which manifest themselves as
an IR cutoff. An example is the holographic principle \cite{holo}, which
states that the field theoretical description overcounts the true
dynamical degrees of freedom and hence, if true, one should impose
extra, non-local constraints on the effective field theory.
Indeed, it is well known that a local quantum field theory
confined to a box of size $\tilde L$ and having a UV cutoff
$\Lambda$ can not fully describe black holes while preserving
unitarity. For any $\Lambda$ there is a sufficiently large volume
for which the entropy of an effective field theory will violate
the Bekenstein bound \cite{bekenstein}; the field theory
overcounts  the true dynamical degrees of freedom. The same
conclusion holds also for quantum gravity: local quantum field
theory is not a good description of gravity because it has too
many degrees of freedom in the UV.

The Bekenstein entropy bound may, however, be satisfied in an
effective field theory if we limit the volume of the system
according to
 \beq \label{be}
\tilde{L}^3\Lambda^3 \leq S_{BH}=\pi \tilde{L}^2M_P^2
 \eeq
where  $S_{BH}$ is the entropy of a black hole of radius
$\tilde{L}$. Thus, the length scale $\tilde{L}$, which provides an
IR cutoff, is determined by the UV cutoff $\Lambda$ and can not be
chosen independently.

As pointed out by Cohen et al. \cite{Cohen:1998zx} (see also
\cite{Thomas:pq}), the actual bound on the volume could be
even stronger. An effective field theory that can saturate
Eq.~(\ref{be}) includes many states with a Schwarzschild radius
much larger than the box size. It seems plausible that we should
exclude such states from the effective field theoretical
description. This lead  Cohen et al. to propose an additional
constraint on the IR cutoff. Since the maximal energy density in
the effective theory is $\Lambda^4$, requiring the energy in a
given volume not to exceed the energy of a black hole of same size
results in the constraint \cite{Cohen:1998zx}
 \beq \label{constr}
\tilde{L}^3\Lambda^4\lesssim \tilde{L}M_P^2~.
 \eeq

It is interesting to observe that if one takes $\tilde{L}$ to be the
size of the
observable universe today i.e. the current particle horizon, given
approximately by the Hubble scale $H_0^{-1}$, saturating Eq. (\ref{constr})
one obtains a vacuum energy density of the right order of magnitude,
consistent with observations \cite{Cohen:1998zx,
  Thomas:pq,Hsu:2004ri,Li:2004rb}. This is an interesting observation
since the size of the cosmological constant is one of the great
finetuning problems in cosmology and also lies at the core of the cosmic
coincidence problem.
Unfortunately, it has been shown that the resulting equation of state for dark
energy  does not agree with data \cite{Hsu:2004ri}. However, as suggested by Li
\cite{Li:2004rb} (see also \cite{Huang:2004ai,Huang:2004wt}), it might be more
natural to take the IR cutoff
to be the size of the volume a given observer eventually will
observe. This is the future event horizon $R_H$, given by
 \beq \label{RH1}
R_H = a\int_t^\infty \frac {dt}{a}~.
 \eeq
Thus, a local field theory should describe only the degrees of
freedom that can ever be observed by a given observer. In this
respect the observer is always in a privileged position. Note
that the future event horizon depends on the equation of state.
This approach could also be in accordance with the black hole
complementarity conjecture \cite{Susskind:1993if} (for a discussion of
de Sitter complementarity see
\cite{Banks:2000fe}), although the way
holography manifests itself is likely to be more convoluted
\cite{Banks:2001px}.
However, for phenomenological purposes it is interesting to assume
that the same IR cutoff that is responsible for the dark energy viewed
as a quantum fluctuation whose size is determined
by the IR/UV duality, is also present in the spectrum of CMB
perturbations. Thus, in effect we assume that the same nonlocal,
possibly holographic theory which at the quantum level ensures
that there is no overcounting of degrees of freedom, at the
classical level is also relevant for the CMB anisotropies. Also,
by virtue of Occam's razor, we could claim that if we observe an
IR cutoff in the CMB, the most economic assumption is that it is
the same IR cutoff which is responsible for the small effective cosmological
constant or dark energy component.

Hence our starting point is the following. Let $\rho_{\Lambda}$ be
the quantum zero-point energy density rendered finite by the UV
cutoff. The total energy in a region of spatial size $\tilde{L}$
should not exceed the mass of a black hole of the same size, or
$\frac{4\pi}{3}\tilde{L}^3\rho_{\Lambda}\leq 4\pi\tilde{L}M_P^2$. The
largest region
$\tilde{L}$ allowed is the one saturating this inequality so that
we may write
 \beq \label{an}
\rho_{\Lambda}=3M_P^2 \tilde{L}^{-2}=3 c^2 M_P^2 R_H^{-2} ~.
 \eeq
Following Li \cite{Li:2004rb} we have introduced the constant $c$, but here
its physical meaning is slightly different since here we have related
the IR cutoff $\tilde{L}$ and the future event horizon $R_H$ by
$R_H\equiv c\tilde{L}$. They should be of same order but could
differ by some factor which in principle would be calculable in
the full theory. Hence we expect that $c\sim {\cal O}(1)$. For
instance, in a generalization of Holographic dark
energy to curved space \cite{Huang:2004ai} it has been proposed that $\tilde L={a}
\sin(y)/{\sqrt{k}}$ where $y=\sqrt{k}R_h/a$. This
would effectively lead to
$c\approx 1/\cos(\sqrt{k}R_h)>1$ \cite{Huang:2004ai}. However, here we
restrict ourselves to flat space and consider $c$ as a free parameter.

To translate $\tilde L$ into an IR cutoff at physical wavelengths we
can simply consider quantization in a spherical potential well with
infinitely high potential walls. The radial
solutions are spherical Bessel functions with
the ground state $j_0$ being proportional to $\sin(kr)/kr$ with $r<r_a$. Here
$r_a$ is the radius on the spherical potential. Demanding that the
solution vanishes at $r=r_a$ one finds that the wavelength of the
ground state is $\lambda_c = 2 r_a$. Thus the IR cutoff $\tilde L$
translates into a cutoff at physical wavelengths given by $\lambda_c =
2\tilde L$.

Using the Friedman equation it can be shown that for a cosmic two
component fluid comprising of a matter component and a
dark energy component $\Omega_{\Lambda}$, in a flat universe
Eq. (\ref{an}) implies
that
 \beq \nonumber
R_H =
a^{3/2}c\frac{1}{\sqrt{\Omega_m^0}H_0}\left(\frac{1-\Omega_{\Lambda}}{\Omega_{\Lambda}}\right)^{1/2}~.
 \eeq
which means that, assuming a flat universe, at present time we have
 \beq \label{RH}
R_H=\frac{c}{\sqrt{\Omega_{\Lambda}^{0}}}H_0^{-1}~.
 \eeq
Let us first point out that
in the CMB angular spectrum the IR cutoff would show up in the
Sachs-Wolfe effect
 \beq\nonumber
C_l=\pi\int_{0}^{\infty}\frac{dk}{k}j_l^2\left(k\eta_0\right)\delta_H^2(k)~,
 \eeq
which is an integral over all comoving
momentum modes. The IR cutoff would result in a lower bound on the
comoving momentum and also turn the integral into a sum over the
discrete momentum modes, hence suppressing the Sachs-Wolfe contribution and CMB
spectrum at small $l$s, so symbolically we can then write the Sachs-Wolfe effect as
 \beq\label{sw}
C_l\simeq N\sum_{k>k_c}\frac{1}{k}j_l^2\left(k\eta_0\right)\delta_H^2(k)~.
 \eeq
Here $k_c$ is the IR comoving momentum mode cutoff,
$\delta_H^2(k)\equiv \frac{2}{25}\mathcal{P_R}(k)$ is the
primordial
  power spectrum, $N$ a normalization factor, and $j_l$ is the spherical Bessel function. If
  the spectrum is a single power law with a spectral index $n$, we have
  $\delta_H^2(k)\propto k^{n-1}$. Note that the spacing in the sum in
  Eq. (\ref{sw}) is not equidistant and the discrete spectrum approaches  rapidly the
  continuum.

This type of cutoff looks much less like a step function than the
type of phenomenological cutoffs associated with inflaton dynamics
discussed in the literature \cite{Contaldi:2003zv,dynacutoff}.
Thus we expect that the IR cutoff due to UV/IR duality must lie at
larger length scales than the usual best fit. In addition, because
of the discreteness, our approach will give rise to oscillatory
features in the low $l$ spectrum. Here it may be noted that the
WMAP team actually considered a toy model with a discrete
equidistant spectrum and suggested that it might fit the data
better than the conventional power spectrum \cite{WMAP}. We should
also mention that since the setup here is different, we do not
expect the geometric patterns that usually appear in models of
finite universes realized as multi-connected spaces
\cite{Levin:2001fg}. In Fig. 1 we show the Sachs-Wolfe effect for
a model with a large scale suppression due to an early epoch with
a blue spectrum, compared to the type of cutoff suggested here.

\begin{figure}[!hbtp] \label{cutoff_step}
\begin{center}
\includegraphics[width=5cm]{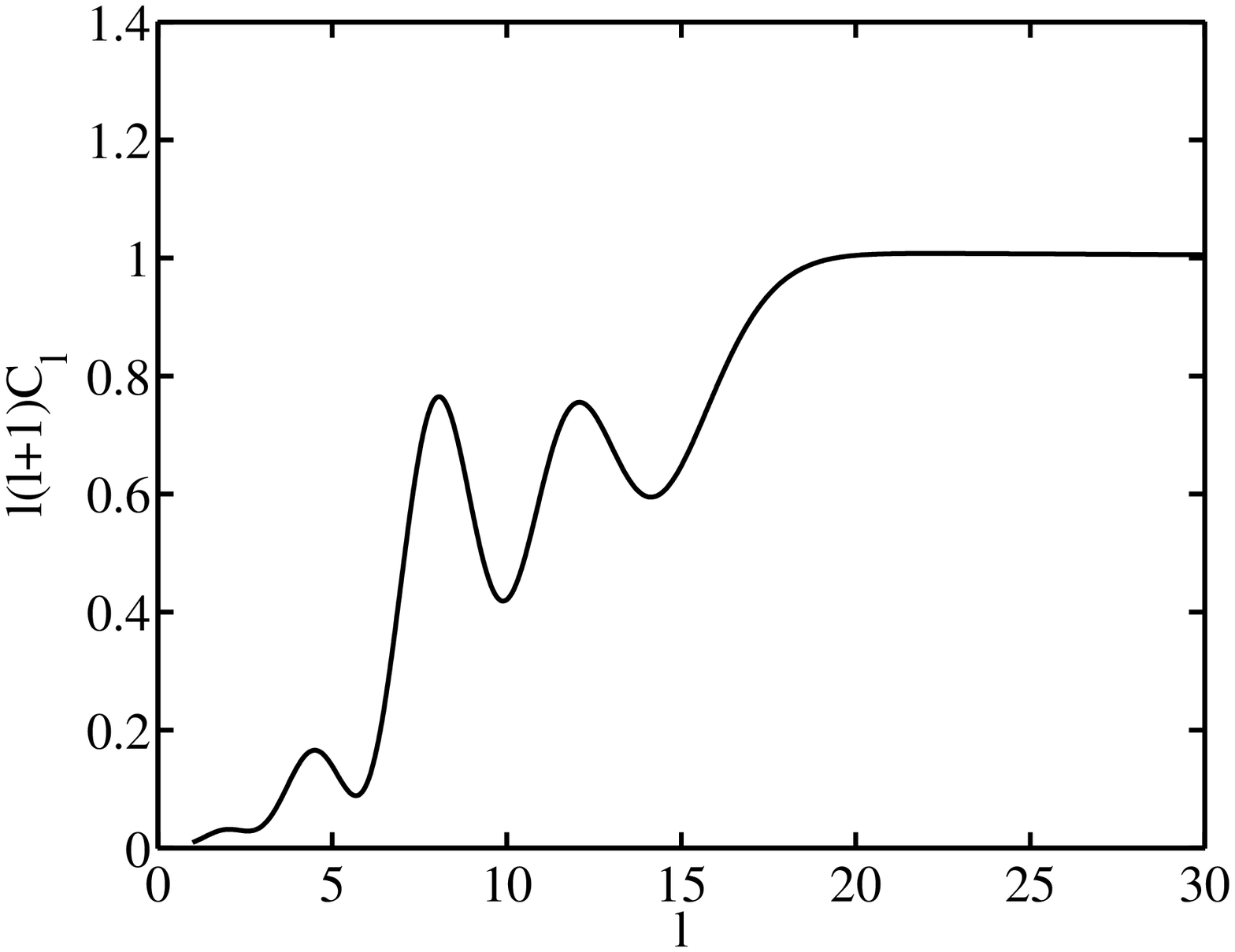}
\includegraphics[width=5cm]{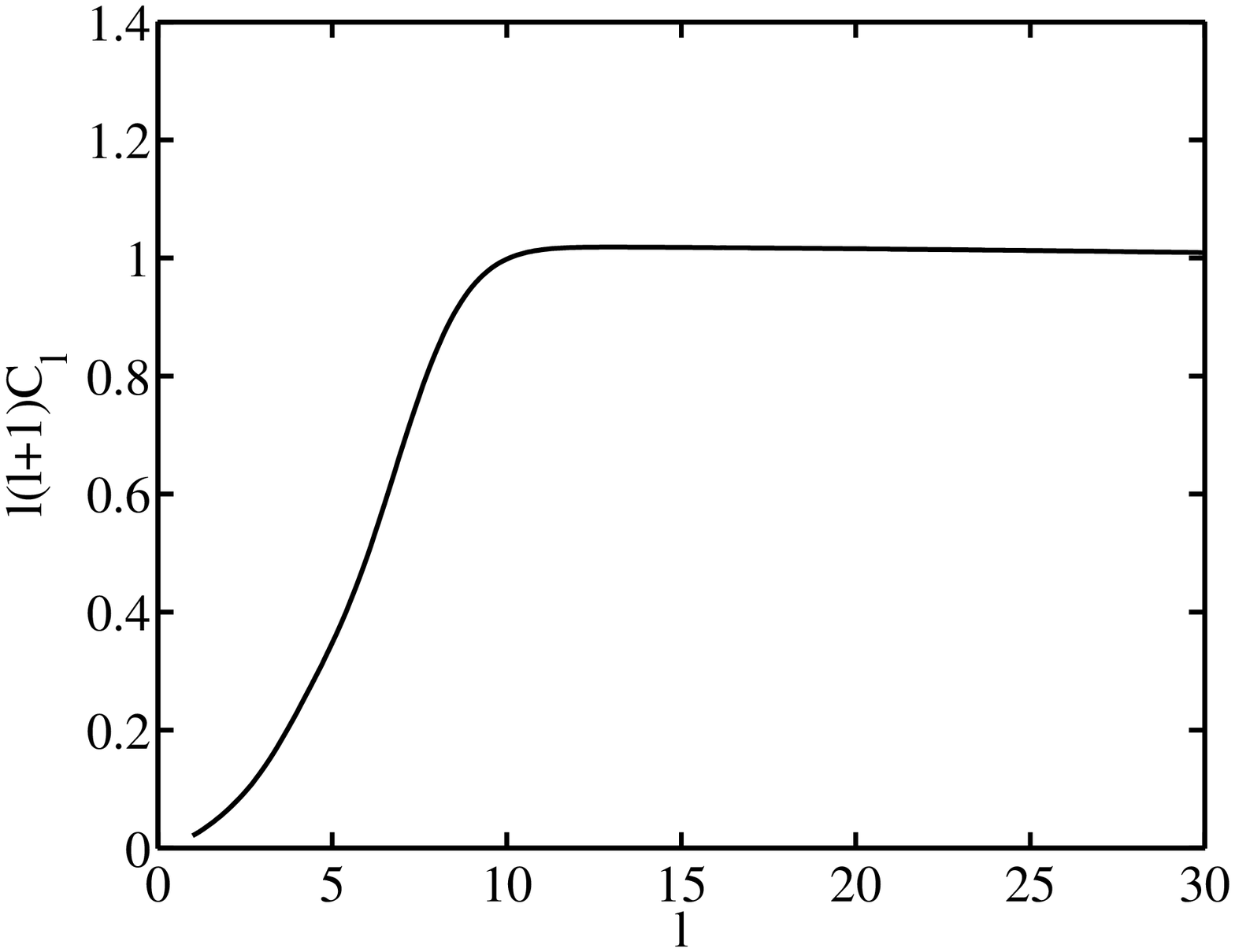}
\end{center}
\caption{Upper panel: the Sachs-Wolfe effect  for a
  spectrum with a sharp cutoff  at $l=10$ due to UV/IR duality. Lower panel:
the Sachs-Wolfe effect for a spectrum with a blue $n=4$ tilt
  for $l<10$ and a flat $n=1$ spectrum for $l>10$.}
\end{figure}

In flat space the multipole number $l$ is given by
$l=k_l(\eta_0-\eta_*)$ where $\eta_0-\eta_*$ is the comoving
distance to the last-scattering surface and $k_l$ is the
corresponding comoving wave number. The comoving distance to last
scattering follows from the definition of comoving time 
\beq\label{eqeta} 
\eta_0-\eta_* =\int_0^{z_*}dz'\frac{1}{H(z')}~. 
\eeq
Assuming that the dominant components of energy in the universe
are dark energy and matter, then 
\bea\nonumber
H^2(z)=H_0^2\left[\Omega_{\Lambda}^0(1+z)^{(3+3w_{\Lambda})}+(1-\Omega_{\Lambda}^0)(1+z)^3\right]~,
\eea 
where $w_{\Lambda}$ is given by the equation of state of dark energy
$p_{\Lambda}=w_{\Lambda}\rho_{\Lambda}$. If $w_{\Lambda}$ is approximated by a
constant, we can find an analytical solution for Eq.~(\ref{eqeta})
in terms of the Gauss hypergeometric function
 \bea\label{eta0}
\eta_0-\eta_*\simeq\frac{2}{\sqrt{\Omega_m^0}H_0}{}_2F_1\left(\frac{1}{2},-\frac{1}{6w_{\Lambda}};1-
\frac{1}{6w_{\Lambda}};-\frac{\Omega_{\Lambda}^0}{\Omega_m^0}\right)
 \eea
In Fig. 2 we show a plot of the solution to Eq.~(\ref{eta0}).
However, one should note that we have slightly overestimated
$\eta_0-\eta_*$ by treating $w_{\Lambda}$ as a constant. An exact numerical
solution is also shown in Fig. 2.

On the other hand, the equation
of state of dark energy can be shown to be related to the
constant $c$ by $w_{\Lambda}^0 = -\frac{1}{3}-\frac{2}{3c}\sqrt{\Omega_{\Lambda}^0}$ \cite{Li:2004rb}.
Thus, the distance to last scattering is a function of $w_{\Lambda}$ which we
can fit by fixing the free parameter $c$. However, from Eq.~(\ref{RH})
we see that $\tilde L$ does not depend on $c$. Using $\Omega_{\Lambda}^0=0.75$ and $\Omega_m^0=0.25$ we find
$\tilde{L}=1.2\times H_0^{-1}$ today, which implies $\lambda_c=2\tilde
L=2.4\times H_0^{-1}$ or $1/k_c\equiv \lambda_c/(2\pi)=1.2/\pi\times H_0^{-1}$.
Since the scale corresponding to the first multipole in the CMB
spectrum is the distance to last scattering, we see that the position
of a large scale cutoff $\tilde L$ relative to the first multipole
has a weak dependence on the equation of state of the dark energy.

We are
interested in the case where the cutoff is in the fiducial part of
the spectrum. For the suppression of the low multipole to be statistically
significant, we probably need to suppress at least the two first
multipoles in the CMB spectrum corresponding to $l=2,3$, i.e.
$1/k_c \lesssim 1/k_{l=3}=(\eta_0-\eta_*)/3$.
Using $1/k_c=1.2/\pi\times H_0^{-1}$ and Eq.~(\ref{eta0}), we find that
the cutoff is indeed in the interesting range\footnote{Even if we took
  the largest possible box, by using in the Sachs-Wolfe effect the
  infrared cutoff on comoving length scales $\tilde L^c=\pi/k_c$
  evaluated at the time of last scattering $c\tilde L^c
  =R_H(t_{LS})/a_{LS}$, the position of the corresponding cutoff $l_c$
  would still be in an interesting range $1.3<l_c<2.6$. It is easy to
  see from Eq. (\ref{RH1}) and (\ref{eqeta}), that the difference in
  the comoving cutoff evaluated today and at the time of last
  scattering is simply $(\eta_0-\eta_*)/c$}.
\begin{figure}[!hbtp] \label{hor2}
\begin{center}
\includegraphics[width=4.8cm]{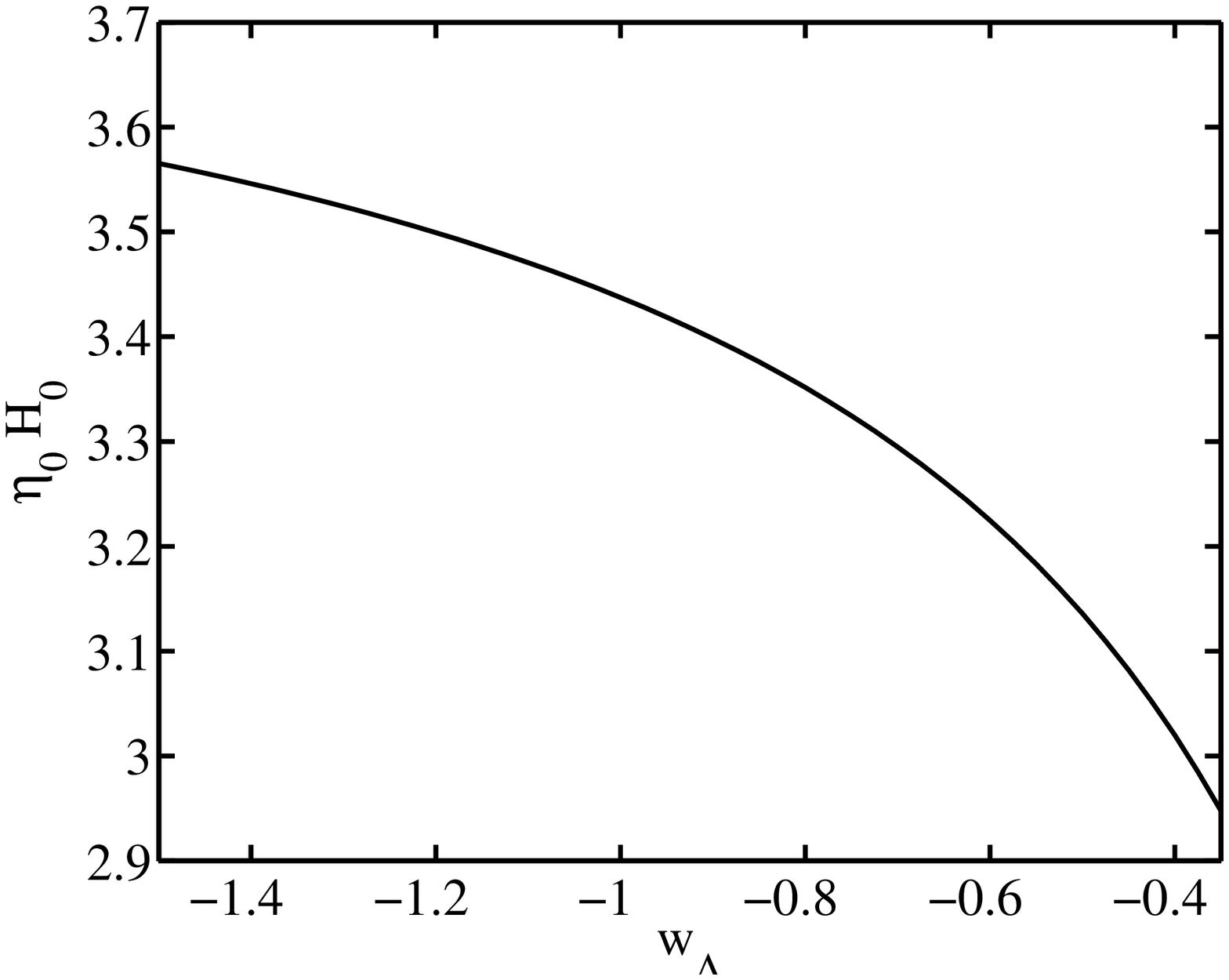}
\includegraphics[width=4.8cm]{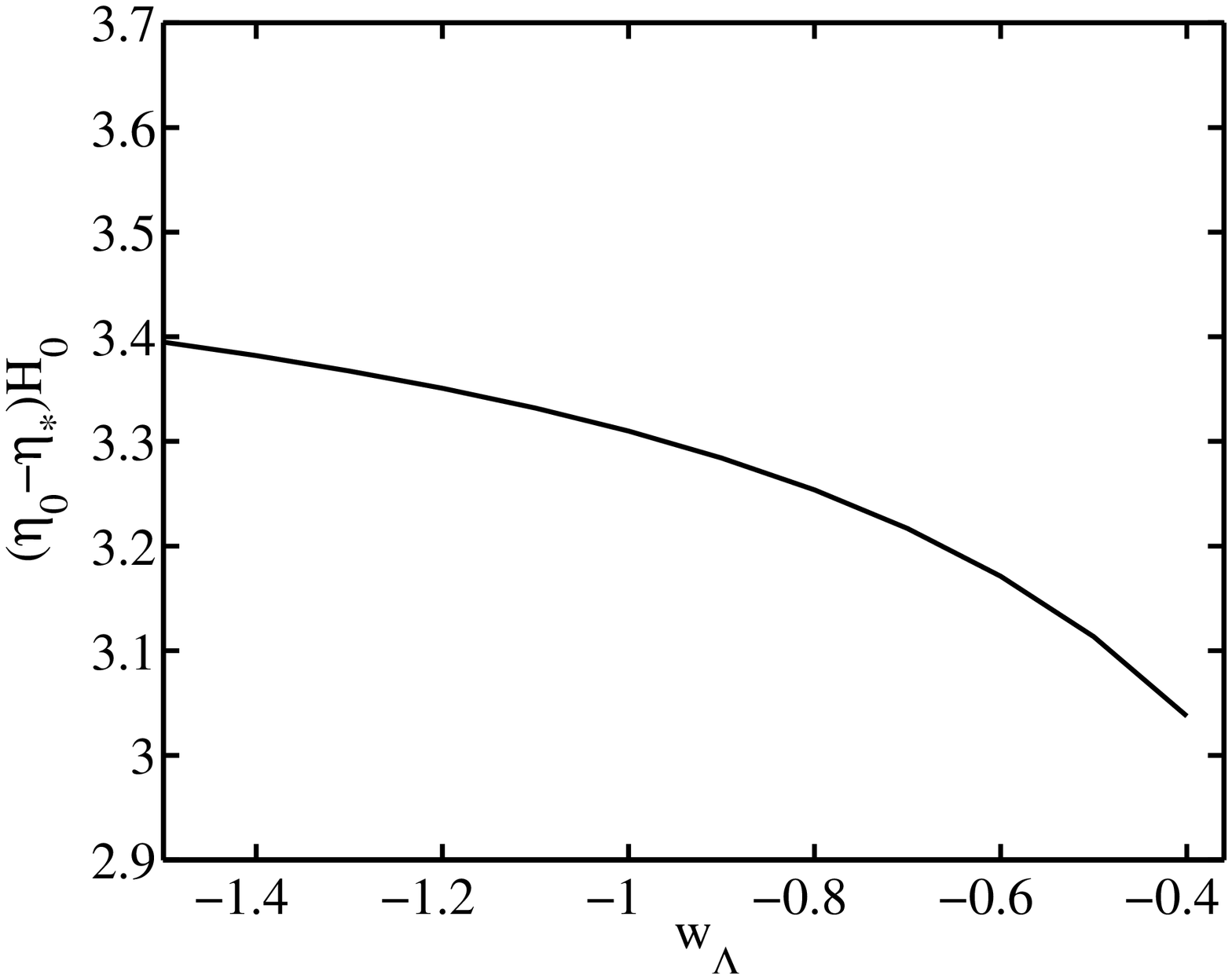}
\end{center}
\caption{Upper panel: The present distance to last scattering,
  $\eta_0$, as a function of the equation of state
  today $w_{\Lambda}^0$, given by Eq.~(\ref{eta0}). Lower panel: A
  numerical solution where the proper redshift dependence of
  $w_{\Lambda}$ has been taken into account (see also Fig. 3).}
\end{figure}
Combining data from WMAP and other cosmic microwave background
experiments with large scale structure, supernova and Hubble Space
Telescope data, the dark energy equation of state has been found
\cite{Melchiorri:2002ux} to be limited by  $-1.38< w_{\Lambda} <-0.82$ at
the 95\% confidence level. Given these experimental limits, the
position of the cutoff $l_c$ in the multipole space falls in the
interval $8.5<l_c<9$. This is probably a bit larger than the
preferred value \cite{Contaldi:2003zv} which corresponds to
$l_c\approx 7$, but still is comfortably within the $2\sigma$
limit corresponding to $l_c\approx 10$; however, these numbers are
just suggestive as a global fit would be needed to find the exact
position and confidence levels of $l_c$.

\begin{figure}[!hbtp] \label{wred}
\begin{center}
\includegraphics[width=4.8cm]{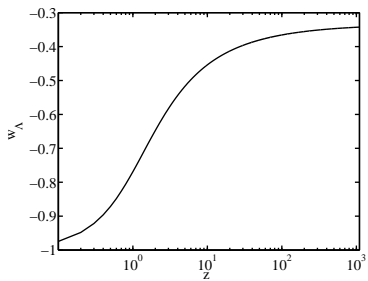}
\end{center}
\caption{The equation of state of dark energy $w_{\Lambda}$
  as a function of redshift $z$ for the specific case of
  $w_{\Lambda}^0=-1$ and $c= 0.87$.}
\end{figure}
The scenario presented
here can be viewed as a toy model of cosmic duality where the IR cutoff plays
an important role. It is worth mentioning that in a universe dominated by
dark energy we actually live inside a finite box,
the so-called "causal diamond" of the static de Sitter coordinates,
which is bounded by the past and future event horizons \cite{Banks:2000fe}. In
cosmic coordinates the finiteness could manifest itself as an
effective IR regulator
of the same order of magnitude as the  future event horizon, which in a
pure de Sitter space determines also the magnitude of the effective
cosmological constant.
Of course, even if an IR/UV duality
is at work in the theory at some fundamental level, the IR
regulator might not be simply related to the future event
horizon but there might
still be a (complicated) relation between the dark energy and the
IR cutoff of the CMB perturbation modes.

In passing, it is interesting to note that in \cite{Nemanja} it is
shown that the future event horizon also limits the number of
e-foldings of inflation that will ever be observable in the CMB spectrum.

We conclude that a relation between the location of the cutoff in
the CMB spectrum and the equation of state of dark energy is
something that can be searched for in the CMB data. However, a
detailed global fit to data, which would simultaneously fix the
value of the cutoff $l_c$ and all the cosmological parameters, is
beyond the scope of this paper, where we merely point out to a
possible interesting connection. For the purposes of falsifying
the cosmic IR/UV duality proposed here, the redshift dependence of
$w_{\Lambda}$ (see Fig 3.), the exact location of the cutoff $l_c$ and the predicted
shape of the spectrum at low $l$ would all be crucial. Here $w_{\Lambda}(z)$
is not an independent quantity but depends on the CMB IR cutoff
$l_c$ and the amount of dark energy. As the time evolution of the
equation of state of dark energy may be detectable too in the near
future, the present scheme can also be put to test.

\begin{acknowledgments}

We would like to thank Nemanja Kaloper, Esko
Keski-Vakkuri and Archil Kobakhidze for interesting comments and
discussions. K.E. is partly supported by the Academy of Finland
grants 75065 and 205800.

\end{acknowledgments}


\begin{thebibliography}{99}


\bibitem{WMAP}
D.~N.~Spergel {\it et al.},
Astrophys.\ J.\ Suppl.\  {\bf 148}, 175 (2003)


\bibitem{Contaldi:2003zv}
C.~R.~Contaldi, M.~Peloso, L.~Kofman and A.~Linde,
JCAP {\bf 0307}, 002 (2003)


\bibitem{holo}
G.~'t Hooft,
arXiv:gr-qc/9310026.
L.~Susskind,
J.\ Math.\ Phys.\  {\bf 36}, 6377 (1995)
R.~Bousso,
Rev.\ Mod.\ Phys.\  {\bf 74}, 825 (2002)



\bibitem{bekenstein}
Phys.\ Rev.\ D {\bf 23}, 287 (1981).
J.~D.~Bekenstein,
Phys.\ Rev.\ D {\bf 7} (1973) 2333.

\bibitem{Cohen:1998zx}
A.~G.~Cohen, D.~B.~Kaplan and A.~E.~Nelson,
Phys.\ Rev.\ Lett.\  {\bf 82}, 4971 (1999)


\bibitem{Thomas:pq}
S.~Thomas,
Phys.\ Rev.\ Lett.\  {\bf 89}, 081301 (2002).

\bibitem{Hsu:2004ri}
S.~D.~H.~Hsu,
arXiv:hep-th/0403052.

\bibitem{Li:2004rb}
M.~Li,
arXiv:hep-th/0403127.

\bibitem{Huang:2004ai}
Q.~G.~Huang and M.~Li,
arXiv:astro-ph/0404229.

\bibitem{Huang:2004wt}
Q.~G.~Huang and Y.~G.~Gong,
arXiv:astro-ph/0403590.
Y.~Gong,
arXiv:hep-th/0404030.
B.~Wang, E.~Abdalla and R.~K.~Su,
arXiv:hep-th/0404057.
R.~Horvat,
arXiv:astro-ph/0404204.
M.~Ito,
arXiv:hep-th/0405281


\bibitem{Susskind:1993if}
L.~Susskind, L.~Thorlacius and J.~Uglum,
Phys.\ Rev.\ D {\bf 48}, 3743 (1993)
L.~Susskind,
Phys.\ Rev.\ Lett.\  {\bf 71}, 2367 (1993)
L.~Susskind and L.~Thorlacius,
Phys.\ Rev.\ D {\bf 49}, 966 (1994)
L.~Susskind and J.~Uglum,
Nucl.\ Phys.\ Proc.\ Suppl.\  {\bf 45BC}, 115 (1996)



\bibitem{Banks:2000fe}
T.~Banks,
arXiv:hep-th/0007146.
R.~Bousso,
JHEP {\bf 0011}, 038 (2000)
T.~Banks and W.~Fischler,
arXiv:hep-th/0102077.
L.~Dyson, M.~Kleban and L.~Susskind,
JHEP {\bf 0210}, 011 (2002)
M.~K.~Parikh, I.~Savonije and E.~Verlinde,
Phys.\ Rev.\ D {\bf 67}, 064005 (2003)
N.~Kaloper, M.~Kleban, A.~Lawrence, S.~Shenker and L.~Susskind,
JHEP {\bf 0211}, 037 (2002)
T.~Banks, W.~Fischler and S.~Paban,
JHEP {\bf 0212}, 062 (2002)
U.~H.~Danielsson, D.~Domert and M.~Olsson,
Phys.\ Rev.\ D {\bf 68}, 083508 (2003)
A.~Albrecht and L.~Sorbo
arXiv:hep-th/0405270.


\bibitem{Banks:2001px}

T.~Banks and W.~Fischler,
arXiv:hep-th/0111142.
C.~J.~Hogan,
Phys.\ Rev.\ D {\bf 66}, 023521 (2002)
F.~Larsen, J.~P.~van der Schaar and R.~G.~Leigh,
JHEP {\bf 0204}, 047 (2002)
A.~Albrecht, N.~Kaloper and Y.~S.~Song,
arXiv:hep-th/0211221.
U.~H.~Danielsson,
JCAP {\bf 0303}, 002 (2003)
E.~Keski-Vakkuri and M.~S.~Sloth,
JCAP {\bf 0308}, 001 (2003)
J.~P.~van der Schaar,
JHEP {\bf 0401}, 070 (2004)
T.~Banks and W.~Fischler,
arXiv:hep-th/0310288.
C.~J.~Hogan,
arXiv:astro-ph/0310532.
F.~Larsen and R.~McNees,
arXiv:hep-th/0402050.


\bibitem{dynacutoff}
J.~M.~Cline, P.~Crotty and J.~Lesgourgues,
JCAP {\bf 0309}, 010 (2003)
B.~Feng and X.~Zhang,
Phys.\ Lett.\ B {\bf 570}, 145 (2003)
M.~Kawasaki and F.~Takahashi,
Phys.\ Lett.\ B {\bf 570}, 151 (2003)
T.~Moroi and T.~Takahashi,
Phys.\ Rev.\ Lett.\  {\bf 92}, 091301 (2004)
Q.~G.~Huang and M.~Li,
JCAP {\bf 0311}, 001 (2003)
X.~J.~Bi, B.~Feng and X.~m.~Zhang,
arXiv:hep-ph/0309195.
Y.~S.~Piao, B.~Feng and X.~m.~Zhang,
Phys.\ Rev.\ D {\bf 63}, 084520 (2000)
Q.~G.~Huang and M.~Li,
arXiv:astro-ph/0311378.
Y.~S.~Piao, S.~Tsujikawa and X.~m.~Zhang,
arXiv:hep-th/0312139.
T.~Multamaki and O.~Elgaroy,
arXiv:astro-ph/0312534.
M.~Liguori, S.~Matarrese, M.~A.~Musso and A.~Riotto,
arXiv:astro-ph/0405544.


\bibitem{Levin:2001fg}
J.~Levin,
Phys.\ Rept.\  {\bf 365}, 251 (2002)


\bibitem{Melchiorri:2002ux}
A.~Melchiorri, L.~Mersini, C.~J.~Odman and M.~Trodden,
Phys.\ Rev.\ D {\bf 68}, 043509 (2003)

\bibitem{Nemanja}
N.~Kaloper, M.~Kleban and L.~Sorbo,
arXiv:astro-ph/406099.

\end{thebibliography}
\end{document}